\documentclass[sigplan,screen]{acmart}

\AtBeginDocument{%
  }

\copyrightyear{2024} 
\acmYear{2024} 
\setcopyright{acmlicensed}\acmConference[ASPLOS '24]{29th ACM International Conference on Architectural Support for Programming Languages and Operating Systems, Volume 1}{April 27-May 1, 2024}{La Jolla, CA, USA}
\acmBooktitle{28th ACM International Conference on Architectural Support for Programming Languages and Operating Systems, Volume 1 (ASPLOS '24), April 27-May 1, 2024, La Jolla, CA, USA}
\acmPrice{15.00}
\acmDOI{10.1145/3617232.3624874}
\acmISBN{979-8-4007-0372-0/24/04}

\usepackage[normalem]{ulem}
\usepackage{graphicx}
\usepackage{textcomp}
\usepackage{enumitem}
\usepackage{booktabs}
\usepackage{multirow}
\usepackage{multicol}
\usepackage{balance}
\usepackage{microtype}
\usepackage{tcolorbox}
\usepackage{makecell}
\usepackage{pifont}
\usepackage[frozencache,cachedir=minted-cache]{minted} 
\usepackage{xcolor}

\newcommand{\tmark}{{\scriptsize\ding{228}}}
\usepackage{amsthm}
\usepackage{amsmath}
\newtheorem{definition}{Definition}
\usepackage{subcaption}
\DeclareCaptionSubType{listing}
\usepackage{csquotes}
\usepackage{mathtools}
\usepackage{setspace}

\DeclareMathAlphabet{\mathcal}{OMS}{cmsy}{m}{n}

\usepackage{titlesec}
\titleformat{\subsubsection}
  {\normalfont\normalsize\bfseries}{\thesubsubsection}{1em}{}

\usepackage[linesnumbered,ruled,vlined]{algorithm2e}

\SetCommentSty{mycommfont}

\usepackage{xspace}

\newcommand{\etal}{\hbox{\emph{et al.}}\xspace}
\newcommand{\eg}{\hbox{\emph{e.g.}}\xspace}
\newcommand{\ie}{\hbox{\emph{i.e.}}\xspace}

\newcommand{\etc}{\hbox{\emph{etc.}}\xspace}

\colorlet{tablerowcolor}{gray!15} 

\colorlet{tablecellcolor}{green!20} 

\setminted{numbersep=5pt}

\newcommand{\thetool}{\textsc{UBfuzz}\xspace}
\newcommand{\asan}{ASan\xspace}
\newcommand{\ubsan}{UBSan\xspace}
\newcommand{\msan}{MSan\xspace}

\newcommand{\para}[1]{\smallskip\noindent\textbf{#1}\xspace}

\graphicspath{ {./figures_small/}}

\newcommand{\revision}[1]{{#1}}

\begin{document}

\title{
\textsc{UBfuzz}: Finding Bugs in Sanitizer Implementations
}

\author{Shaohua Li}
\email{shaohua.li@inf.ethz.ch}

\affiliation{%
  \institution{ETH Zurich}
  \country{Switzerland}
}

\author{Zhendong Su}
\email{zhendong.su@inf.ethz.ch}

\affiliation{%
  \institution{ETH Zurich}
  \country{Switzerland}
}

\begin{abstract}

In this paper, we propose a testing framework for validating sanitizer implementations in compilers.
Our core components are (1) a program generator specifically designed for producing programs containing undefined behavior (UB), and (2) a novel test oracle for sanitizer testing.
The program generator employs Shadow Statement Insertion, a general and effective approach for introducing UB into a valid seed program. The generated UB programs are subsequently utilized for differential testing of multiple sanitizer implementations. Nevertheless, discrepant sanitizer reports may stem from either compiler optimization or sanitizer bugs. To accurately determine if a discrepancy is caused by sanitizer bugs, we introduce a new test oracle called \emph{crash-site mapping}.

We have incorporated our techniques into \thetool, a practical tool for testing sanitizers. Over a five-month testing period, \thetool successfully found 31 bugs in both GCC and LLVM sanitizers. These bugs reveal the serious false negative problems in sanitizers, where certain UBs in programs went unreported.
This research paves the way for further investigation in this crucial area of study.


\end{abstract}

\ccsdesc[500]{Software and its engineering~Compilers}
\ccsdesc[500]{Security and privacy~Software and application security}

\keywords{Undefined Behavior, Sanitizer, Compiler, Program Generation, Fuzzing}
  
\maketitle

\section{Introduction}\label{sec:intro}
Undefined behaviors (UB), such as buffer overflow, integer overflow, \etc, are often responsible for creating security weaknesses in software~\cite{manes2019art,wang2012undefined}. 
Sanitizers are crucial in enabling the large-scale detection of security vulnerabilities caused by UB~\cite{serebryany2016continuous,serebryany2016sanitize}. Popular sanitizers include Address Sanitizer (\asan)~\cite{serebryany2012addresssanitizer} for memory access errors, Undefined Behavior Sanitizer (\ubsan)~\cite{llvm2023ubsan} for various undefined behaviors, and Memory Sanitizer (\msan)~\cite{stepanov2015memorysanitizer} for uninitialized memory uses.
Technically, sanitizers are integrated into compilers. When a sanitizer is enabled, various checks are inserted into a program during compilation. If a check is violated at run-time, an error is reported.
Owing to their superior capability and usability, sanitizers have assisted developers in discovering numerous critical vulnerabilities. For instance, by fuzzing with sanitizers, the Google OSS-Fuzz project has reported over 20K UBs in hundreds of open-source projects~\cite{ding2021empirical,ossfuzz}.
While substantial research and engineering efforts have been made toward devising efficient fuzzers~\cite{li2023accelerating,nagy2021same} and reducing sanitizer costs~\cite{zhang2022debloating,zhang2021sanrazor,jeon2020fuzzan}, the robustness and reliability of sanitizers --- essential for detection effectiveness --- have received little attention from both academia and industry.



Both GCC and LLVM, the two most popular C/C++ compilers, support sanitizers.
Over the past five years, there were only 29 bug reports related to sanitizer correctness in the bug trackers of GCC and LLVM. Most of these reports (66\%) were false positive issues, where sanitizers did not miss UBs but instead incorrectly reported correct executions as containing UB.
False positive issues are indeed easy to be noticed in practice. For example, typical compiler testing work~\cite{yang2011finding,le2014compiler,livinskii2020random} involves generating valid programs as input, which can be trivially adapted to identify false positive issues.
Conversely, false negative bugs in sanitizers typically result in a UB being missed and are thus difficult to be observed.






Figure~\ref{lst:gcc-asan} illustrates a code snippet that triggers a false negative bug in GCC \asan.
Since \texttt{d} points to the starting location of \texttt{b[2]}, the dereference \texttt{*(d+k)} at line 8 will cause a stack-buffer-overflow.
When we compile and run this code with GCC \asan at -O0, \asan crashes the execution and generates a report as expected (Figure~\ref{lst:gcc-asan} top right). However, at -O2, it unexpectedly misses this UB (Figure~\ref{lst:gcc-asan} bottom right). 
This is a false negative bug of GCC \asan.
As a UB detection tool, false negative bugs in sanitizers lead to missing UBs, thereby significantly impeding their effectiveness.

In this paper, we aim to detect false negative (FN) bugs in sanitizers.
Despite its criticality and importance, to the best of our knowledge, there exists no work that has systematically investigated this problem. 
We introduce the first effective testing framework for finding FN bugs in sanitizers. 
At a high level, the general testing workflow is (1) generating a UB program, \ie, a program exhibiting undefined behavior, and (2) compiling it with sanitizers and executing the compiled binary. If no sanitizer report on a UB program is produced, a potential sanitizer bug is detected.
Two main challenges exist, which we will discuss next.

\begin{figure}[tp]
    \captionsetup[subfigure]{aboveskip=1pt}
\vspace{16pt}
    \begin{subfigure}[t]{.23\textwidth}
\begin{minted}[linenos,xleftmargin=1.5em,fontsize=\small]{C}
struct a { int x };
struct a b[2];
struct a *c=b, *d=b;
int k = 0;
int main() {
  *c = *b;
  k = 2;
  *c = *(d+k);
  return c->x;
}
\end{minted}
    \end{subfigure}
\hfill
    \begin{subfigure}[t]{.24\textwidth}

\begin{minted}[escapeinside=@@,fontsize=\scriptsize]{C}
(command line)
 @\textbf{\$ gcc -O0 -fsanitize=address a.c}@
 @\textbf{\$ ./a.out}@
 ==1==ERROR: AddressSanitizer: @\emph{}@
 @\emph{stack-buffer-overflow}@ in a.c:8
 @\textbf{\$}@


(command line)
 @\textbf{\$ gcc -O2 -fsanitize=address a.c}@
 @\textbf{\$ ./a.out}@
 @\textbf{\$}@
\end{minted}
    \end{subfigure}
    \vspace{-6pt}
\caption[cap]{Line 8 in \texttt{a.c} contains a stack-buffer-overflow (left). GCC \asan at -O0 successfully detects it (top right). GCC's Asan at -O2, however, overlooks it (bottom right). \footnotemark}

    \label{lst:gcc-asan}
\end{figure}
\footnotetext{{\url{https://gcc.gnu.org/bugzilla/show_bug.cgi?id=105714}}}

\medskip\noindent\textbf{Challenge 1: } \textit{UB program generation.}\\
To detect FN bugs, abundant and diverse UB programs should be available. The automated generation of valid programs for compiler testing has been extensively researched. 
Tools like Csmith~\cite{yang2011finding} can generate a wide variety of valid C programs that are free from UB. 
However, the generation of programs exhibiting various types of UB, which is essential for sanitizer testing, remains unexplored. 
For instance, to test \asan, programs with memory safety bugs such as buffer-overflow, use-after-free, use-after-scope, \etc, are needed.
One might consider randomly mutating a valid program, for example, deleting statements or altering variable values, to introduce UBs. As we will demonstrate in our evaluation~\S\ref{sec:eval-generator}, this naive mutation-based method is ineffective in generating UB programs---most of the mutated programs do not have UB. Furthermore, the generated programs encompass only a few UB types and are unable to find any FN bugs in sanitizers.

\medskip\noindent\textbf{Our solution: } \textit{Shadow Statement Insertion.}\\
We propose a general approach for introducing UB into a valid program. Given a program and a target UB such as buffer overflow, our approach first applies static and dynamic analysis to learn the program's runtime state and identify a specific program location where the target UB can be introduced. Subsequently, we insert a new statement into the program such that the chosen program location triggers a UB. We term our approach \emph{shadow statement insertion}. 
For instance, the original program of Figure~\ref{lst:gcc-asan} does not have line 7 and is thus free of UB. To introduce a buffer overflow, our tool analyzes the code and identifies that the pointer \mintinline{c}|d| points to the stack buffer \mintinline{c}|b| of size 8 bytes. It then inserts \mintinline{c}|k=2| to overflow the buffer access at line 8.
As will be detailed in~\S\ref{sec:approach}, following the same framework, our design can be generalized to other UB types.

\medskip\noindent\textbf{Challenge 2: } \textit{Compiler optimization significantly complicates sanitizer testing.}\\
Given an input UB program, a natural approach is to examine whether a compiler's sanitizer such as \texttt{gcc -O2 -fsanitize} \texttt{=address} can detect the UB. If no report is produced, one might assume that ``a sanitizer FN bug is discovered''. However, this is not true due to compiler optimizations.

\begin{figure}[t]
\centering
\vspace{6pt}
\includegraphics[width=0.48\textwidth]{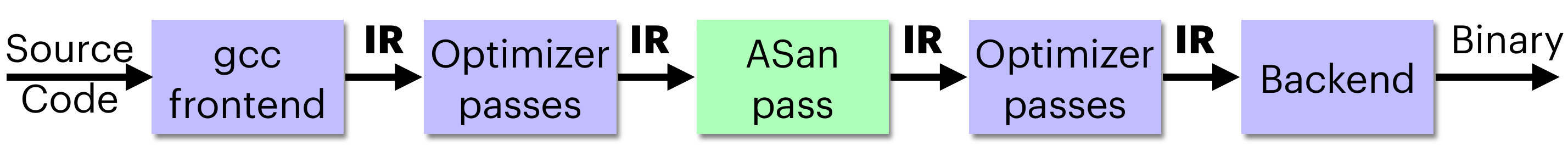}
\caption{The high level compilation pipeline of \asan in GCC/gcc.}
\label{fig:san-workflow}
\vspace{-10pt}
\end{figure}

\begin{figure}[t]
\centering
\includegraphics[width=0.49\textwidth]{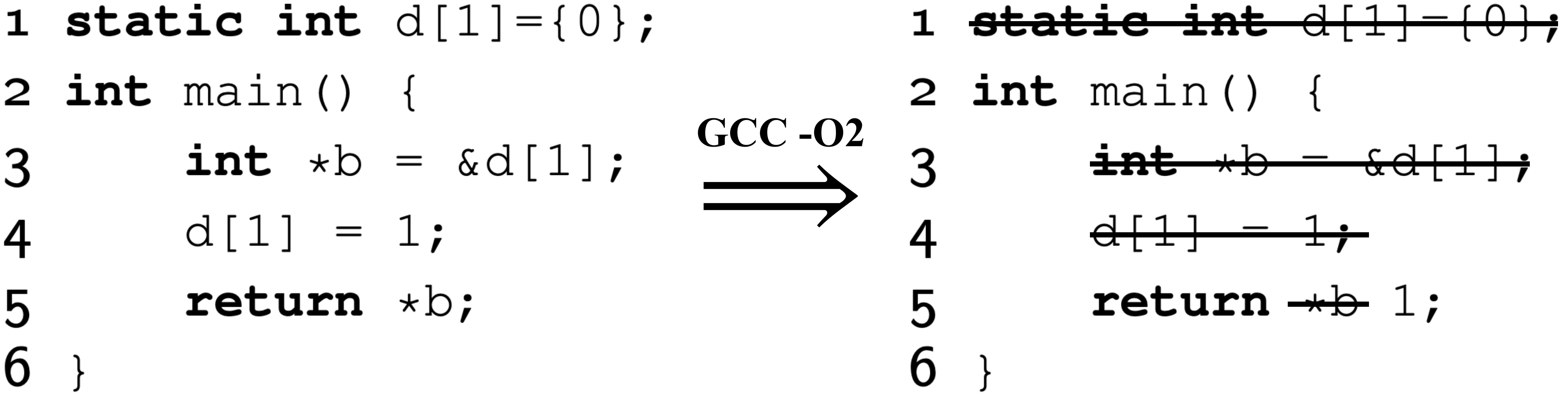}
\caption{GCC -O2 optimizes away the UB code, thus \asan cannot discover the UB.}
\vspace{-10pt}
\label{fig:intro-not-fn}
\end{figure}

Sanitizers are implemented as passes in compilers' pipeline. Figure~\ref{fig:san-workflow} illustrates the high-level pipeline in GCC compilation with \asan enabled. The \asan pass collaborates with other optimizer passes to compile a program. 
Previous research~\cite{li2023finding,Isemann2023look,wang2013towards} has shown that compiler optimizers always presume that the input program does not contain UB, resulting in the elimination of certain UBs by optimizer passes.
Figure~\ref{fig:intro-not-fn} provides an example where both \mintinline{c}|d[1]=1| at line 4 and \mintinline{c}|*b| at line 5 trigger stack-buffer-overflow UB. Nonetheless, if we compile it with \asan at -O2 
 (\mintinline{c}|gcc -O2| \mintinline{c}|-fsanitize=address|), no UB report will be produced. 
\emph{Different from the previous example in Figure~\ref{lst:gcc-asan}, this is not a sanitizer bug.} 
The reason is that early optimization passes at GCC -O2 optimize away all the UB code, as depicted on the right side of Figure~\ref{fig:intro-not-fn}. Since there is no UB present in the input IR to the \asan pass, \asan cannot uncover the UB in the source code.
As there is no UB in the final compiled binary, \asan is not considered buggy in this example.

A natural follow-up question is \emph{can we only consider unoptimized compilers such as with -O0?}
The answer is no for two main reasons. First, even with -O0, some basic optimizations, such as constant folding, may still optimize away the UB code.
Second, many sanitizer FN bugs only exist at higher optimization levels as demonstrated in Figure~\ref{lst:gcc-asan}. Testing sanitizers only at -O0 may fail to detect many critical FN bugs.

Similarly, differential testing across different compilers is ineffective as it is impossible to determine whether a discrepant report is caused by a sanitizer FN bug or merely due to compiler optimizations. For instance, although GCC \asan at \texttt{-O0} and \texttt{-O2} produce different results in both Figure~\ref{lst:gcc-asan} and \ref{fig:intro-not-fn}, the latter is caused by compiler optimizations.

\medskip\noindent\textbf{Our solution: } \textit{Crash-site mapping as the test oracle.}\\
We introduce a novel test oracle, \emph{crash-site mapping}, to accurately discern whether discrepant sanitizer reports stem from sanitizer FN bugs or compiler optimizations.
Given two binaries, $b_c$ and $b_n$, compiled by two compilers such as GCC \asan at \texttt{-O0} and \texttt{-O2}, executing $b_c$ results in a crash while executing $b_n$ exits normally. Here, the crash of $b_c$ means that the sanitizer successfully reports the UB while the normal exit of $b_n$ means that the sanitizer does not report the UB.
Our primary approach involves using a debugger to trace the execution of both binaries. If the crash location in $b_c$ is also executed by $b_n$, we can infer that the compiler does not eliminate the UB, and thus, it is highly probable that a sanitizer FN bug is present. A more detailed example will be provided in Section~\ref{sec:illustrative}. 
As our evaluation will demonstrate, our crash-site mapping oracle can effectively identify discrepancies caused by compiler optimizations.

We realized our solutions in a tool named \thetool. It can automatically generate a substantial number of UB programs and accurately identify sanitizer FN bugs.
The example we showcased in Figure~\ref{lst:gcc-asan} was found by \thetool. We reported this FN bug to the GCC team, who confirmed and fixed it. The root cause is that in some cases, GCC \asan would ``forget'' to insert checks to specific memory accesses, thus resulting in missed UB reports. 
During a five-month testing period, \thetool uncovered a total of 31 new FN bugs in \asan, \ubsan, and \msan from both GCC and LLVM. 
\revision{
We open-sourced our implementation to facilitate future research\footnote{\url{https://github.com/shao-hua-li/UBGen}}.
}

In summary, we make the following contributions:
\begin{itemize}[leftmargin=15pt, topsep=5pt]
\setlength\itemsep{0.2em}
    \item We introduce a general approach, \emph{Shadow Statement Insertion}, for generating UB programs.
    \item We design \emph{crash-site mapping} as an effective oracle for sanitizer testing.
    \item Based on the proposed UB generator and test oracle, we develop the automated tool \thetool for testing sanitizers.
    \item We report our extensive evaluation of \thetool, which successfully identified 31 sanitizer FN bugs.
\end{itemize}


\section{Illustrative Examples}\label{sec:illustrative}
This section illustrates (1) how \thetool generates UB programs for a target UB, and (2) how our crash-site mapping test oracle works.

\subsection{UB Program Generation}\label{sec:illustrative-generation}

\definecolor{color1}{HTML}{2171b5}
\newcommand{\colorLOG}[1]{\textcolor{color1}{#1}}
\definecolor{color2}{HTML}{fc8d59}
\newcommand{\colorMUT}[1]{\textcolor{color2}{#1}}
\definecolor{color3}{HTML}{238443}
\newcommand{\colorINSERT}[1]{\textcolor{color3}{#1}}
\newcommand{\circlednumber}[1]{\raisebox{.5pt}{\textcircled{\raisebox{-.9pt} {#1}}}}

\begin{figure}
    \begin{minted}[linenos,xleftmargin=2em,escapeinside=@@]{c}
struct a { int x };
struct a b[2];
struct a *c = b, *d = b;
int k = 0;
int main() {
    \end{minted}
    \begin{minted}[linenos=false,xleftmargin=2em,escapeinside=@@]{c}
  @\colorLOG{\textbf{LOG\_BufRange}(&b[0], sizeof(b));\circlednumber{1}}@
    \end{minted}
    \begin{minted}[linenos,firstnumber=6,xleftmargin=2em,escapeinside=@@]{c}
  *c = *b;
    \end{minted}
    \begin{minted}[linenos=false,xleftmargin=2em,escapeinside=@@]{c}
    
  @\colorINSERT{\textbf{k = 2;} \circlednumber{3}}@
  @\colorMUT{\textbf{LOG\_BufAccess}(d+k);\circlednumber{2}}@
    \end{minted}
    \begin{minted}[linenos,firstnumber=7,xleftmargin=2em,escapeinside=@@]{c}
  *c = *(d+k);
  return c->x;
}
    \end{minted}
\caption{Code instrumentation for UB insertion.}
\label{lst:example-gen}
\vspace{-10pt}
\end{figure}

The uncolored (black) code in Figure~\ref{lst:example-gen} is the seed program. We now demonstrate how we mutate this seed program to the UB program in Figure~\ref{lst:gcc-asan}, with the goal of introducing a \emph{stack-buffer-overflow} UB. Our generator works as follows:
\begin{enumerate}[label={\textbf{Step~\arabic*.}}, wide, labelwidth=!, labelindent=0pt, itemsep=3pt, topsep=2pt]
    \item Insert profiling statements for all stack buffers. 
    Since there is only one global stack buffer \texttt{b[2]} in this seed program, a single profiling statement is inserted as indicated by \colorLOG{\circlednumber{1}}. 
    Next, identify all code constructs with the potential to exhibit the target UB.
    Given that our target UB is stack-buffer-overflow, we statically locate all memory accesses. All the pointer dereferences \texttt{*b}, \texttt{*c}, and \texttt{*(d+k)} at lines 6 and 7 are eligible. 
    For the simplicity of presentation, we only show the profiling statement \colorMUT{\circlednumber{2}} for \texttt{*(d+k)}.
    \item After the instrumentation, we compile and execute the code to obtain its runtime information including the memory range of buffer \texttt{b} and the memory address of \texttt{d+k}.
    \item To introduce a stack-buffer-overflow for \texttt{*(d+k)} at line 7, we mutate the value of \texttt{k} such that it overflows the pointed-to buffer. 
    Since we have learned that the buffer \texttt{b} has a size of 8 bytes and \texttt{d} points to the starting location of \texttt{b}, we insert the shadow statement \texttt{k=2} \colorINSERT{\circlednumber{3}} to introduce a stack-buffer-overflow at line 7. 
    One might question whether setting \texttt{k} to an arbitrarily large enough value would also introduce a buffer overflow. However,
    due to the design limitation of \asan, it can only detect overflows of up to 32 bytes. Consequently, 
    the valid range of overflowed addresses falls between 8 $\sim$ 32 bytes beyond \texttt{b}. Our precise runtime analysis enables us to precisely mutate \texttt{d+k} such that it falls within this range.
\end{enumerate}

Finally, all logging statements \colorLOG{\circlednumber{1}}\colorMUT{\circlednumber{2}} are removed while the shadow statement \colorINSERT{\circlednumber{3}} is kept. The resulting UB program is identical to the one presented in Figure~\ref{lst:gcc-asan}.

\begin{figure}[t]
\captionsetup[subfigure]{aboveskip=1pt}
\vspace{20pt}
\begin{subfigure}[b]{.23\textwidth}
    \begin{minted}[fontsize=\small]{gas}
...             #line,offset
andl  %r8d, %esi #10,8
movq  %rax, %rdi #10,8
callq 0x10e0     #10,8
    \end{minted}
\caption{The last three executed instructions in $b_c$.}
\label{lst:gcc-O0-example1.out}
\end{subfigure}
\hfill
\begin{subfigure}[b]{.24\textwidth}
    \begin{minted}[fontsize=\small]{gas}
movq  ptr(%rip),%rax #10,3
movl  $0xfff,(%rax)  #10,8
    \end{minted}
\caption{The executed instructions that are from line 10 in $b_n$.}
\label{lst:gcc-O2-example1.out}
\end{subfigure}
\caption{Partial executed instructions in $b_c$ and $b_n$. The comment shows the corresponding line and offset in the source code.}
\end{figure}

\newcommand{\greentext}[1]{\fboxsep=2.0pt\colorbox{gray!20}{#1}}
\newcommand{\colorRedTable}[1]{\greentext{#1}}
\newcommand{\stmt}[2]{{\texttt{Stmt\{}#1\texttt{\}{#2}}}}
\newcommand{\ifstmt}[2]{{\texttt{if(}#1\texttt{){#2}}}}
\newcommand{\whilestmt}[2]{{\texttt{while(}#1\texttt{){#2}}}}
\newcommand{\stmtshadow}[2]{{\texttt{$\Delta$(}#1\texttt{){#2}}}}
\newcommand{\arraysize}[1]{{\footnotesize\textsc{ArraySize}}(#1)}
\newcommand{\bufferrange}[1]{{\footnotesize\textsc{BufferRange}}(#1)}
\newcommand{\scope}[1]{{\footnotesize\textsc{Scope}}(#1)}
\begin{table*}[t]
    \centering
    \small
    \caption{UB conditions and shadow statements. 
    The first three columns describe the conditions for certain code constructs to not have the target UB. 
    The fourth column demonstrates the location where our shadow statements will be inserted.
    The fifth column presents the effect of each shadow statement. 
    The last column lists the instantiation of each shadow statement in our implementation.
    Here, $x,\widehat{x},y,\widehat{y}, \widehat{c}$ are ($n+1$)-bit integers; $p,q$ are pointers; $a$ is an array with capacity \arraysize{$a$}; $lhs\xrightarrow{val}rhs$ represents the value of $lhs$ is $rhs$.
    }
    \renewcommand{\arraystretch}{2.8}
    \begin{tabular}{l|ll||lll}
    \toprule
    UB & \makecell[l]{Code\\Constrcut} & \makecell[l]{Sufficient condition for\\not having the UB} & \makecell[l]{Shadow \\Statement \textbf\texttt{$\Delta$(${\cdot}$)}} & \makecell[l]{Effect of \textbf\texttt{$\Delta$(${\cdot}$)}} & Instantiation\\\midrule
    
\makecell[l]{Buf. Overflow\\(Array)} 
& {$a[x]$}{} & $0 \leq x < \mbox{\arraysize{$a$}}$ 
& \makecell[l]{\colorRedTable{\stmtshadow{$x$}{;}}\\ \stmt{$a[x]$}{;}} 
& \makecell[l]{
$x\xrightarrow{val}v$ and $(v < 0\ \lor $\\
$v \geq \mbox{\arraysize{$a$}})$
}
& \makecell[l]{\colorRedTable{$\widehat{x}=v-x;$}\\ \stmt{$a[x+\widehat{x}]$}{;}}\\

\makecell[l]{Buf. Overflow\\(Pointer)} 
& {$*p$}{} & $p \in \mbox{\bufferrange{$p$}}$ 
& \makecell[l]{\colorRedTable{\stmtshadow{$p$}{;}}\\ \stmt{$*p$}{;}} 
& \makecell[l]{
$p\xrightarrow{val} q$ and \\
$q \notin \mbox{\bufferrange{$p$}}$ 
}
& \makecell[l]{\colorRedTable{$\widehat{c}=q-p;$}\\ \stmt{$*(p+\widehat{c})$}{;}}\\

Use After Free 
& {$*p$}{} & $\forall free(q)$, !alias($p, q$) 
& \makecell[l]{\colorRedTable{\stmtshadow{$p$}{;}}\\ \stmt{$*p$}{;}} 
& \makecell[l]{
$p\xrightarrow{val} q$ and $q$ is freed
}
& \makecell[l]{\colorRedTable{$free(p);$}\\ \stmt{$*p$}{;}}\\

Use After Scope 
& {$*p$}{} & \makecell[l]{\scope{$*p$} $\in$ \scope{$p$}}
& \makecell[l]{\colorRedTable{\stmtshadow{$p$}{;}}\\ \stmt{$*p$}{;}} 
& \makecell[l]{
$p\xrightarrow{val} q$ and \\
\scope{$*q$} out of \scope{$p$} 
}
& \makecell[l]{\colorRedTable{$p=q;$}\\ \stmt{$*p$}{;}}\\

Null Ptr. Deref.
& {$*p$}{} & $p\neq$\textsc{Null}
& \makecell[l]{\colorRedTable{\stmtshadow{$p$}{;}}\\ \stmt{$*p$}{;}} 
& \makecell[l]{
$p\xrightarrow{val}$\textsc{Null} 
}
& \makecell[l]{\colorRedTable{$p=0;$}\\ \stmt{$*p$}{;}}\\

Integer Overflow 
& {$x$ op $y$}{} & $x$ op $y \in [-2^{n}, 2^{n}-1]$
& \makecell[l]{\colorRedTable{\stmtshadow{$x, y$}{;}}\\ \stmt{$x$ op $y$}{;}} 
& \makecell[l]{
$x\xrightarrow{val}v_0, y\xrightarrow{val}v_1$ and \\
$v_0$ op $v_1 \notin [-2^{n}, 2^{n}-1]$ 
}
& \makecell[l]{\colorRedTable{$\widehat{x}=v_0-x, \widehat{y}=v_1-y;$}\\ \stmt{$(x+\widehat{x})$ op$(y+\widehat{y})$}{}}\\

Shift Overflow 
& \makecell[l]{{$x \ll y$}{} or \\ {$x \gg y$}{}} & $ 0 \leq  y < n$ 
& \makecell[l]{\colorRedTable{\stmtshadow{$y$}{;}}\\ \stmt{$x \ll y$}{;}} 
& \makecell[l]{
$y\xrightarrow{val}v$ and \\
$v < 0 \lor  v \geq n$ 
}
& \makecell[l]{\colorRedTable{$\widehat{y}=v-y;$}\\ \stmt{$x \ll (y+\widehat{y})$}{;}}\\

Divide by Zero 
& \makecell[l]{{$x / y$}{} or \\ {$x \% y$}{} } & $y\neq0$
& \makecell[l]{\colorRedTable{\stmtshadow{$y$}{;}}\\ \stmt{$x / y$}{;}}
& \makecell[l]{
$y\xrightarrow{val}0$
}
& \makecell[l]{\colorRedTable{$\widehat{y}=-y;$}\\ \stmt{$x / (y+\widehat{y})$}{;}}\\

\makecell[l]{Use of Uninit. \\Memory}
& \makecell[l]{\ifstmt{$x$}{} or \\ \whilestmt{$x$}{} } & $x$ is uninitialized
& \makecell[l]{\colorRedTable{\stmtshadow{$x$}{;}}\\ \stmt{$x$}{;}}
& \makecell[l]{
$x\xrightarrow{val}$ uninit. memory
}
& \makecell[l]{
\colorRedTable{\textbf\texttt{int}~ $\widehat{x};$}\\ 
\stmt{$x +\widehat{x}$}{;}
}\\

    \bottomrule
    \end{tabular}
    \vspace{10pt}
    \label{tab:ub_condition}
\end{table*}

\subsection{Crash-Site Mapping as the Test Oracle}
After a UB program is generated, we use at least two compilers with sanitizer enabled to compile it. We then execute the compiled binaries to examine whether there is a discrepant report. If one of the binaries crashes while the other does not, we need to determine if the discrepancy is caused by compiler optimizations. We refer to the crashing binary as $b_{c}$ and the non-crashing binary as $b_{n}$. For the code snippet in Figure~\ref{lst:example-gen}, $b_{c}$ is compiled by GCC \asan at -O0, while $b_n$ is from -O2.
Our crash-site mapping works as follows:
\begin{enumerate}[label={\textbf{Step~\arabic*.}}, wide, labelwidth=!, labelindent=0pt, itemsep=3pt, topsep=2pt]
    \item \emph{Analyze $b_{c}$:} We utilize a debugger\footnote{In our implementation, we use LLDB and its python API to automate our analysis. More details in the evaluation section.} to track the execution of $b_c$ and obtain the last executed site, \ie, the crash-site. Figure~\ref{lst:gcc-O0-example1.out} shows the last three executed instructions of $b_c$, the last of which indicates the crash-site is at (line 10, offset 8). This means that executing the instruction compiled from (line 10, offset 8) in the source code results in a crash, \ie, a sanitizer report.
    \item \emph{Analyze $b_{n}$:} Once again, we use the debugger to track the execution of $b_n$. Since we have learned the crash-site in Step 1, we only need to monitor if the crash-site is also executed in $b_n$. Figure~\ref{lst:gcc-O2-example1.out} shows a part of the execution in $b_n$. We can observe that (line 10, offset 8) is executed as well.
    \item \emph{Mapping:} Since the crash-site from $b_{c}$ is also executed in $b_{n}$, we classify this program as triggering a sanitizer FN bug. Otherwise, the discrepancy would be classified as being caused by compiler optimizations.
\end{enumerate}

For the program shown in Figure~\ref{fig:intro-not-fn}, the crash site from GCC \asan -O0 is not present in GCC \asan -O2, which indicates that the inconsistent sanitizer reports are caused by compiler optimizations. Our evaluation in Section~\ref{sec:eval-oracle} will demonstrate that crash-site mapping can accurately identify discrepancies resulting from compiler optimizations.

\begin{figure}[tp]
\begin{minted}[escapeinside=@@]{C}
 int a[5]; int x=1;       int a[5]; int x=1;
 a[x] = 1;          @$\Longrightarrow$@   x = 5; //@\stmtshadow{$x$}{;}@
                          a[x] = 1;
\end{minted}
\vspace{-12pt}
\caption{The expression $x=5$ is inserted as the shadow statement to introduce a buffer overflow in $a[x]$.}
\label{fig:shadow-example}
\end{figure}

\section{Approach}\label{sec:approach}

We first analyze sufficient conditions for a valid program to be free from UB, which motivates the design of our shadow statement insertion method. Then, we introduce the proposed UB program generation approach.
Finally, we present the crash-site mapping as the test oracle for sanitizer testing.

\subsection{UB Conditions and Shadow Statement}

\noindent\textbf{\emph{Code constructs that are free of UB.}}\quad
To generate UB programs, we need to first understand how UB is triggered. The first three columns in Table~\ref{tab:ub_condition} list the conditions for certain code constructs to \emph{not have} the UB, as specified in the C standard~\cite{c17standard}. For instance, the first shown UB is buffer-overflow. For an array access $a[x]$ to be free from this UB, the index $x$ should be positive and less than the array size. 
Another example is signed integer overflow. As long as the calculation of $x$\texttt{ op }$y$ falls within the range of $[-2^n,2^n-1]$, it is free from this UB.
We can conclude that for a code construct that has the adventure of a UB, as long as the given condition is met, it is free from the UB.

\medskip\noindent\textbf{\emph{Code constructs that have UB.}}\quad
Since we now understand the conditions for having UBs in certain code constructs, to introduce a UB, we can simply find a way to break the condition. In this paper, we utilize \emph{shadow statements} to achieve this purpose.
For a valid program $\mathcal{P}$ that contains a code construct $expr$, we introduce a UB by placing a shadow statement \greentext{\stmtshadow{$expr$}{}} before the code construct as follows:
\begin{displayquote}
\vspace{-3pt}
  \greentext{\stmtshadow{$expr$}{;}}\\
  \stmt{$expr$}{;}
\vspace{-3pt}
\end{displayquote}
\noindent
The shadow statement \stmtshadow{$expr$}{} is designed to change the evaluation value of $expr$ such that when executing \stmt{$expr$}{}, the UB condition is triggered. The fourth and fifth columns in Table~\ref{tab:ub_condition} list the shadow statements and their effects.
For example, to introduce a buffer overflow to $a[x]$, the inserted shadow statement \stmtshadow{$x$}{} changes the value of $x$ to $v$, which is out of the range of array $a$. Figure~\ref{fig:shadow-example} illustrates a concrete example where the shadow statement $x=5$ is inserted before the array access.

\noindent
There are two key questions. 
The first is how to understand the target effect of \stmt{$expr$}{}. In the above example, we need to know the concrete range of array $a$, which our generator uses dynamic analysis to obtain.
The second is how to instantiate the shadow statement. Once we know the target effect of \stmt{$expr$}{}, there are plenty of ways to instantiate it. In the above example, we can also choose $x=x+4$ or $x = x*4+1$ as the shadow statement, which results in the same effect as $x=5$.
The last column in Table~\ref{tab:ub_condition} lists the instantiations we used in our implementation. Details will be discussed next.

\subsection{UB Program Generator}

\begin{algorithm}[tp]
\SetFuncSty{textsf}
\setstretch{1.2}
\DontPrintSemicolon
\SetKwInput{KwInput}{Input}                
\SetKwInput{KwOutput}{Output}              
\SetKwFunction{Profile}{Profile}
\SetKwFunction{GetMatchedExpr}{GetMatchedExpr}
\SetKwFunction{ShadowStmt}{SynShadowStmt}
\SetKwFunction{Mutate}{Insert}
\SetKwFunction{Append}{append}

\SetKwFunction{FGen}{Generator}
\SetKwFunction{FSum}{Sum}
\SetKwFunction{FSub}{Sub}

\SetKwProg{Proc}{procedure}{:}{\KwRet}
\Proc{\FGen{Program $\mathcal{P}$, Input $\mathcal{I}$, UBType $\mathcal{U}$}}{
    \tcp{find all matched $expr$ to a given UB}
    $E \gets \GetMatchedExpr(\mathcal{P},\ \mathcal{U})$\;
    $\widehat{\mathit{prof}} \gets \Profile(\mathcal{P},\ \mathcal{I},\ \mathcal{U},\ E)$  \tcp*[r]{profiling}
    $P_{UB} \gets [\ ]$\; 
    \ForEach{$expr \in E$}{
        \tcp{synthesize a shadow statement}
        $\Delta(expr) \gets \ShadowStmt(expr,\ \widehat{\mathit{prof}},\ \mathcal{U})$\;
        \tcp{insert the shadow statement}
        $\mathcal{P}' \gets \Mutate(\mathcal{P},\ \Delta(expr)\ )$\;
        \tcp{append the new UB program}
        $P_{UB}.\Append(\mathcal{P}')$\;
    }
    \Return{$P_{UB}$}\;
}
\caption{UB program generation}
\label{alg:ub-generator}
\end{algorithm}

Algorithm~\ref{alg:ub-generator} shows the general process of generating UB programs. Given a seed program $\mathcal{P}$ and an associated input $\mathcal{I}$, our goal is to generate UB programs that contain the target UB type $\mathcal{U}$ on the input $\mathcal{I}$. Our generator works as follows:
\begin{enumerate}[label={\textbf{Step~\arabic*.}}, wide, labelwidth=!, labelindent=0pt, itemsep=3pt, topsep=2pt]
    \item \emph{Expression Matching} (line 2): find all expressions in $\mathcal{P}$ that have the target code constructs for the given UB. For example, given buffer overflow, according to Table~\ref{tab:ub_condition}, this procedure will find all array accesses \ie, $a[x]$, and pointer dereferences \ie, $*p$.
    \item \emph{Program Profiling} (line 3): instrument and run $\mathcal{P}$ on the input $\mathcal{I}$ to collect an execution profile that contains the required runtime information such as the allocated buffers and pointer addresses.
    \item \emph{Shadow statement synthesis and insertion} (lines 6-7): for each target $expr$, query the execution profile to synthesize a shadow statement, and insert it into the seed program to obtain a UB program.
\end{enumerate}

As indicated by the algorithm, our generator has the following features:
\begin{itemize}
    \item \emph{Target UB type needs to be specified when being invoked.} For every invocation, our generator will generate a set of UB programs that all have the same target UB type.
    \item \emph{Only one UB in every generated program.} Lines 6-8 in Algorithm~\ref{alg:ub-generator} show that for every matched expression, the generator generates a UB program with shadow statement insertion. Consequently, for each generated program, there is a single UB.
    \item \emph{Multiple UB programs for one invocation.} The for loop (line 5) signifies that a UB program is generated for each of the matched expressions. Ultimately, the generator returns a set of UB programs, all containing the same UB type.
\end{itemize}
\noindent
Next, we detail each of the above steps.

\subsubsection{Expression Matching --- \textsf{GetMatchedExpr($\cdot$)}}\label{sec:expression-match}
Given a seed program and a target UB, we statically scan the program to find all expressions that match the code constructs as specified in Table~\ref{tab:ub_condition}.
For example, suppose our target UB is signed integer overflow, we will find all expressions that have the form of $x $ \texttt{op} $y$, where \texttt{op} is an arithmetic operator such as $+$, $-$, and $*$.
After scanning, all matched expressions will be saved into $E$, each item in which contains the matched expression and its location in $\mathcal{P} $.

\subsubsection{Program Profiling --- \textsf{Profile($\cdot$)}}\label{sec:profiling}

An execution profile provides runtime information about the program, which is essential for our shadow statement synthesis. We define the execution profile $\widehat{\mathit{\mathit{prof}}}$ as follows.

\begin{definition}[Execution Profile]\label{def:profile}
Given a program $\mathcal{P}$, an input $\mathcal{I}$, and the target expression list ${E}$, the execution profile $\widehat{\mathit{prof}}$ records the following information during running $\mathcal{P}$ with $\mathcal{I}$:  (1) all the values of expressions in ${E}$ observed, and (2) all the allocated and freed stack and heap memory address ranges. 
\end{definition}

To facilitate easy access to the execution profile, let $e$ denote $expr$, we define the following queries to obtain concrete information from $\widehat{\mathit{prof}}$:
\begin{itemize}[itemsep=3pt, topsep=2pt]
\item \textsc{$Q_{liv}$($\widehat{\mathit{prof}}$, $e$)}: return \texttt{true} if $e$ is in the live region; otherwise, return \texttt{false}. This information is inferred by checking if $e$ has a value in $\widehat{\mathit{prof}}$. If it does, then it is located in the live region; otherwise $\widehat{\mathit{prof}}$ is unable to obtain its value.
\item \textsc{$Q_{val}$($\widehat{\mathit{prof}}$, $e$)}: return the value of $e$.
\item \textsc{$Q_{mem}$($\widehat{\mathit{prof}}$, $e$)}: $e$ is a pointer or an array. Return the memory range that $e$ points to. If the memory has already been freed, return \texttt{false}.
\item \textsc{$Q_{scp}$($\widehat{\mathit{prof}}$, $e$)}: return the scope of $e$. We extend $\widehat{\mathit{prof}}$ with scope information obtained from Clang's LibTooling.
\end{itemize}

\revision{
In our implementation, given a new seed program, we first obtain its execution profile $\widehat{\mathit{prof}}$ and then synthesize UB programs. Thus, the profiling overhead for all UB types is identical. This is an implementation choice because when testing sanitizers, \thetool by default generates all the supported UB programs for one seed program.
}


\subsubsection{Shadow Statement Synthesis and Insertion\\ 
--- \textsf{SynShadowStmt($\cdot$)} \& \textsf{Insert($\cdot$)}}
For a target UB, the synthesized shadow statement should have the effect as shown in the fifth column in Table~\ref{tab:ub_condition}. In theory, there are numerous ways to instantiate a shadow statement. 
For instance, as previously illustrated in Figure~\ref{fig:shadow-example}, expressions like $x=5$, $x=x+4$, $x=x*4+1$, and many others, all satisfy the requirement. In our implementation, for each shadow statement, we choose the simplest instantiation to minimize changes to the seed program. The last column in Table~\ref{tab:ub_condition} lists the instantiations. Details are as follows:
\begin{itemize}\setlength{\itemsep}{5pt}
    \item \emph{Buffer overflow (array)}: We introduce an auxilary variable $\widehat{x}$ to the original expression to obtain {$a[x+\hat{x}]$}{}. $\Delta(expr)$ is $\widehat{x}=v-x$. 
    The value of $v$ is obtained by calculating the memory size from \textsc{$Q_{mem}$($\widehat{\mathit{prof}}$, $a$)}; the value of $x$ is obtained via \textsc{$Q_{val}$($\widehat{\mathit{prof}}$, $x$)}.
    This instantiation does not change any other program semantics except for our target expression.
    
    \item \emph{Buffer overflow (pointer)}: Similarly, we first introduce an auxilary variable $\widehat{x}$ to obtain {$*(p+\widehat{x})$}{}. $\Delta(expr)$ is $\widehat{x}=q-p$, where values of $q$ and $p$ are obtained via \textsc{$Q_{mem}$($\widehat{\mathit{prof}}$, $p$)} and \textsc{$Q_{val}$($\widehat{\mathit{prof}}$, $p$)}, respectively.
    
    \item \emph{Use after free}: $\Delta(expr)$ is $free(p)$.
    
    \item \emph{Use after scope}: $\Delta(expr)$ is $p=q$, where \textsc{$Q_{scp}$($\widehat{\mathit{prof}}$, $*q$)} is not within the scope of \textsc{$Q_{scp}$($\widehat{\mathit{prof}}$, $p$)}.
    
    \item \emph{Null pointer dereference}: $\Delta(expr)$ is $p=(void *)0$.
    
    \item \emph{Integer overflow}: We first introduce auxiliary variables $\widehat{x}$ and $\widehat{y}$ to the original expression to obtain {$(x+\widehat{x})$ \texttt{op} $(y+\widehat{y})$}{}. Then the shadow statement is set to $\widehat{x}=v_0-x, \widehat{y}=v_1-y$. The values of $x$ and $y$ are obtained via \textsc{$Q_{val}$($\widehat{\mathit{prof}}$, $x$)} and \textsc{$Q_{val}$($\widehat{\mathit{prof}}$, $y$)}.
    To find the proper values $v_0$ and $v_1$, we adopt Monte Carlo to sample from $[-2^{n}, 2^{n}-1]$ such that $(x+\widehat{x})$ \texttt{op} $(y+\widehat{y})$ exceeds the range of an ($n+1$)-bit integer.
    
    \item \emph{Shift overflow}: We first introduce an auxiliary variable $\widehat{y}$ to obtain {$x\ll(y+\widehat{y})$}{} or {$x\gg(y+\widehat{y})$}{}, and then set the shadow statement to $\widehat{y}=v-y$. The value of $y$ is obtained via \textsc{$Q_{val}$($\widehat{\mathit{prof}}$, $y$)} and $v$ is a random value satisfying $v<0\; \lor \geq n$.
    
    \item \emph{Divide by zero}: We first introduce an auxiliary variable $\widehat{y}$ to obtain {$x / (y+\widehat{y})$}{}, and then set the shadow statement to $\widehat{y}=-y$. The value of $y$ is obtained via \textsc{$Q_{val}$($\widehat{\mathit{prof}}$, $y$)}.
    
    \item \emph{Use of uninitialized memory}: We first introduce an auxiliary variable $\widehat{x}$ to obtain {$x+\widehat{x}$}, and then set $\Delta(expr)$ to $\texttt{int}\ \ \widehat{x}$. Since $\widehat{x}$ is uninitialized, $x+\widehat{x}$ becomes uninitialized as well.
\end{itemize}

\revision{
Some of the above operations, such as \emph{Buffer overflow (pointer)}, need to know the precise pointer information to accurately synthesize shadow statements. Such pointer information can be obtained via \textsc{$Q_{mem}$($\widehat{\mathit{prof}}$, $e$)}, which achieves this goal by logging all allocated pointers' addresses and used pointers (see the example in \S~\ref{sec:illustrative-generation} \textbf{Step 1}). Thus, we do not need any separate pointer analysis.
}



\subsubsection{Discussions}

As the evaluation will demonstrate, our generator can effectively generate interesting UB programs for sanitizer testing. 
Despite its effectiveness, it does also come with certain limitations. 
First, our UB program generator relies on seed programs. If seed programs are not expressive enough, \thetool cannot generate useful UBs. Fortunately, seed program generators like Csmith have proven effective in exercising rich language features~\cite{li2023finding,le2014compiler}.
Second, the list of UB in \thetool is non-exhaustive. The C17 standard~\cite{c17standard} lists 219 UB types.
\revision{
Not all UBs are supported by sanitizers. We selected UBs that are (1) supported by at least one sanitizer and (2) included in the CWE list~\cite{cwe} which enumerates all common weaknesses in C by the MITRE community. Our supported UBs cover all UBs studied by the related work~\cite{wang2013towards, Isemann2023look}.
}
Generally, each UB comes with a root cause and can be represented in a generic pattern, as demonstrated in our approach. For instance, using pointer subtraction to determine size is UB if two pointers point to different objects~\cite{cwe469}. Realizing this UB in \thetool would require knowledge of the address ranges of each object and pointer, which can be easily obtained through dynamic profiling. We chose not to realize this UB because none of the existing sanitizers support its detection.

\revision{
For each seed program, as a new UB is introduced into it, its semantics is consequently altered.
We clarify that preserving the seed program’s semantics is not necessary in our application scenario because we only require the resulting program to contain the desired UB. Second, all UB programs have invalid, often nondeterministic semantics because (1) their semantics rely on how the compiler deals with UBs, and (2) the compiler has full freedom in handling code with UBs.
Nevertheless, UBFuzz still preserves the runtime semantics of a seed program up to the mutation site.
}

\subsection{Crash-site Mapping as the Test Oracle}

\begin{algorithm}[tp]
\SetFuncSty{textsf}
\SetDataSty{textit}
\setstretch{1.1}
\DontPrintSemicolon
\SetKwInput{KwInput}{Input}                
\SetKwInput{KwOutput}{Output}              
\SetKwFunction{Profile}{Profile}
\SetKwFunction{GetMatchedExpr}{GetMatchedExpr}
\SetKwFunction{ShadowStmt}{SynShadowStmt}
\SetKwFunction{Mutate}{Mutate}
\SetKwFunction{Append}{append}
\SetKwFunction{GetExecutedSites}{GetExecutedSites}
\SetKwFunction{Init}{Init}
\SetKwFunction{SetBreakpoint}{SetBreakpoint}
\SetKwFunction{isalive}{IsAlive}
\SetKwFunction{NextInst}{NextInstruction}
\SetKwFunction{Continue}{Continue}

\SetKwData{debugger}{debugger}

\SetKwFunction{Ffilter}{IsBug}
\SetKwFunction{FGetExecutedSites}{GetExecutedSites}

\SetKwProg{Proc}{procedure}{:}{\KwRet}
\Proc{\Ffilter{Binary $b_c$, Binary $b_n$}}{
    $S_c \gets \GetExecutedSites(b_c)$\;
    $S_n \gets \GetExecutedSites(b_n)$\;
    \If{$S_c[-1] \in S_n$}{\
        \Return{True}\;
    }
    \Else{
        \Return{False}\;
    }
}

\Proc{\FGetExecutedSites{Binary $b$}}{
    $S \gets [\ ]$\;
    $\debugger.\Init(b)$\;
    \While{$\debugger.\isalive()$}{
        $l \gets \debugger.curr\_line$\;
        $o \gets \debugger.curr\_offset$\;
        $S.\Append((l, o))$ \;
        $\debugger.\NextInst()$\;
    }
    \Return{$S$}\;
}
\caption{Crash-Site Mapping}
\label{alg:crashsitemapping}
\end{algorithm}

With the generated UB programs, we employ differential testing across multiple compilers to find sanitizer FN bugs.
Without loss of generality, assume that we have two compilers $\mathcal{C}_c$ and $\mathcal{C}_n$ with the same sanitizer enabled, \eg, GCC \asan at -O1 and LLVM \asan at -O1. The corresponding compiled binaries are $b_c$ and $b_n$. 
Suppose that executing $b_c$ results in a crash while $b_n$ exits normally. Here, the crash in $b_c$ means that the sanitizer in $\mathcal{C}_c$ \emph{successfully} reports the UB; the normal exits of $b_n$ means that the sanitizer in $\mathcal{C}_n$ \emph{does not} report any UB.
As analyzed in Section~\ref{sec:intro}, the discrepancy can arise from a sanitizer FN bug or merely compiler optimizations.
Our \emph{crash-site mapping} can identify the true cause of the discrepancy.
Before introducing our approach, we formally define \emph{crash site}.

\begin{definition}[Crash Site]\label{def:crashsite}
A binary $b_i$ is compiled from program $\mathcal{P}$ and running $b_i$ results in a crash. We denote the last executed instruction as $\widehat{inst}$. If $\widehat{inst}$ corresponds to the line $l$ and offset $o$ in $\mathcal{P}$, then the crash site of $b_i$ is (l, o).
\end{definition}

Our key insight is that if the crash site in $b_c$ is also executed by $b_n$, the compiler $\mathcal{C}_n$ does not optimize away the UB-triggering expression in $\mathcal{P}$, thus the discrepancy is caused by a sanitizer FN bug in $\mathcal{C}_n$.
Algorithm~\ref{alg:crashsitemapping} details our approach.

We first obtain the executed sites of both $b_c$ and $b_n$, \ie, all the executed (line, offset) in $\mathcal{P}$ (line 2-3). If the last executed site in $b_c$, \ie, the crash site, is also present in $b_n$'s executed sites, return true (line 4-5). Otherwise, return false (line 7). 
To obtain all executed sites in a binary, we utilize a debugger to track the execution. The procedure \textsf{GetExecutedSites()} provides the necessary steps. Note that when the debugger reaches an instruction, the $\mathit{debugger.curr\_line}$ and $\mathit{debugger.curr\_offset}$ return the line and offset in the source program that the instruction corresponds to.
The effectiveness of crash-site mapping depends on its accuracy in identifying discrepancies caused by compiler optimizations. Our evaluation in Section~\ref{sec:eval-oracle} will show that it can achieve near-perfect accuracy.

\begin{table}[tp]
    \centering
    \footnotesize
    \caption{UB types supported by each sanitizer.}
    \renewcommand{\arraystretch}{1.4}
    \setlength{\tabcolsep}{1pt}
{\renewcommand{\arraystretch}{1.4}
    \begin{tabular}[t]{lc}
    \toprule
    \textbf{UB} & \textbf{Sanitizer} \\ 
    \midrule
\makecell[l]{Buf. Overflow(Array)} 
& \asan, \ubsan\\

\makecell[l]{Buf. Overflow(Pointer)} 
& \asan\\

Use After Free 
& \asan\\

Use After Scope 
& \asan\\

Null Ptr. Deref.
& \ubsan\\

    \bottomrule
    \end{tabular}
}\hfill
\begin{tabular}[t]{lc}
    \toprule
    \textbf{UB} & \textbf{Sanitizer} \\
    \midrule
Integer Overflow 
& \ubsan\\

Shift Overflow 
& \ubsan\\

Divide by Zero 
& \ubsan\\

\makecell[l]{Use of Uninit. \\Memory}
& \msan\\
    \bottomrule
    \end{tabular}
    \label{tab:san-ub}
\end{table}

\section{Empirical Evaluation}
Our evaluation is based on the following research questions:
\begin{enumerate}[label=\textbf{RQ\arabic*}, itemsep=3pt, topsep=2pt]

\item \textbf{Bug-finding:} Is \thetool effective in finding FN bugs in sanitizers?

\item \textbf{UB generator:} How effective is our UB program generator in constructing interesting UB programs?

\item \textbf{Crash-site mapping:} How accurate is the crash-site mapping test oracle in identifying discrepancies caused by compiler optimizations?

\item \textbf{Code coverage:} Can \thetool improve code coverage?

\end{enumerate}


\subsection{Implementation and Evaluation Setup}\label{sec:eval-setup}

\para{Implementation.}
Our realization of \thetool consists of $\sim$2,000 lines of C++ and $\sim$4,400 lines of Python.
We use Clang's LibTooling~\cite{libtooling} to implement expression matching in Section~\ref{sec:expression-match} and program instrumentation for execution profiling in Section~\ref{sec:profiling}. We utilize LLDB~\cite{lldb} as the debugger in crash-site mapping and use its Python API to automate the analysis process.
\revision{
Our \thetool can run in a fully automated manner in testing sanitizers, including UB program generation, crash-site mapping, and debugging procedures. Once launched, our tool will automatically generate UB programs and use the crash-site mapping algorithm to find FN bugs.
}

\para{Compilers and sanitizers.}
Sanitizers are integrated into compilers. We used \thetool to test the latest development versions of both GCC and LLVM, which support the most widely-used sanitizers, namely \asan, \ubsan, and \msan. 
Note that, \msan is not yet supported by GCC.
Since sanitizers are typically used with optimizations, we enabled the most frequently used optimization levels, namely -O0, -O1, -Os, -O2, and -O3, in both compilers for differential testing.

\para{Seed programs.}
We use Csmith~\cite{yang2011finding} --- a random C program generator --- to produce valid seed programs. There are three main reasons:
\begin{enumerate}[label={{(\arabic*)}}, wide, labelwidth=!, labelindent=0pt, itemsep=3pt, topsep=2pt]
    \item Csmith is adopted by a lot of compiler testing work~\cite{wang2023compilation,theodoridis2022finding,le2014compiler} and has become the de facto default program generator in testing C compilers; 
    \item Csmith can generate complex programs with rich features (\eg, pointer and integer operations), thus offering \thetool abundant opportunities to generate diverse UB programs; and 
    \item programs generated by Csmith are self-contained meaning that they do not take inputs and can be executed.
\end{enumerate}

\para{Hardware.}
We conducted all our evaluations on two Linux servers running Ubuntu 20.04 LTS. Both are equipped with an AMD EPYC 7742 64-Core CPU and 256GB RAM.

\para{Testing process.}
Our testing process is fully automated and runs continuously. We first use Csmith to generate a well-formed seed program. Then, for each of the supported UB, we apply \thetool to generate UB programs from the seed. For each of the UB programs, as we know their UB type, we use compilers with the corresponding sanitizer enabled to compile and run it. Table~\ref{tab:san-ub} lists the supported sanitizers for each UB.
Once a discrepancy is found, we apply crash-site mapping to decide if it is a sanitizer FN bug. If so, we use C-Reduce to reduce the UB program and report the reduced program to the respective bug tracker.
During a period of five months, we sporadically tested the sanitizers. \thetool generated around 130 million UB programs. 
\revision{Note that, since our work focuses on in-house testing, we assume no adversary is present. Thus, successful sanitization always results in a crash. }

\subsection{RQ1: Bug Finding}\label{sec:bug-finding}


Table~\ref{tab:bug_status} summarizes the sanitizer bugs we discovered during our testing period. Overall, we reported 31 bugs. The developers have confirmed 20 of them as previously unknown, real bugs. This highlights the significant bug-finding capability of \thetool. Of all these bugs, 6 of them have been fixed and all the fixed bugs are in GCC.
The relatively high number of unfixed bugs could be attributed to the fact that many of the reported bugs are introduced since the launch of sanitizers and affect \emph{all stable compiler versions}. Our later analysis will show this fact. We also experienced that the LLVM developers were less responsive than GCC and mostly only labeled our reports as sanitizer bugs without further diagnosis. We are strict in marking a bug as confirmed --- only if the developers have clearly diagnosed it and responded to us.
This causes although \thetool found nearly the same number of bugs in GCC and LLVM, most confirmed and all fixed bugs are found in GCC.

\begin{figure}[tp]
\centering
\begin{minipage}[b]{.23\textwidth}
  \centering
    \includegraphics[width=\linewidth]{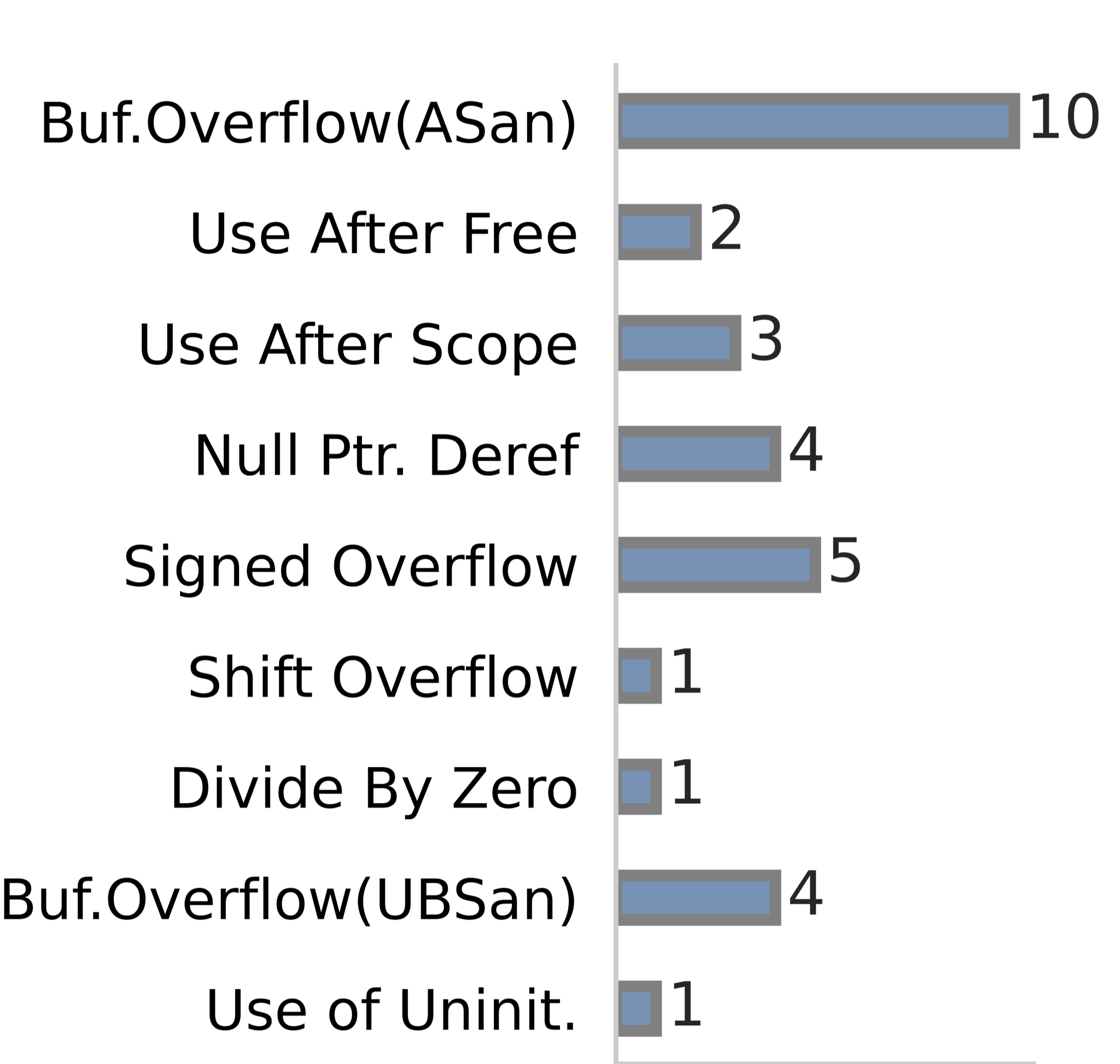}
    \caption{Number of bugs triggered by each kind of UB.}
    \label{fig:bugs_each_ub}
\end{minipage}%
\hfill
\begin{minipage}[b]{.23\textwidth}
  \begin{minted}[linenos,xleftmargin=3em,fontsize=\small]{C}
int a, b;
int main() {
 int *s = &a;
 for(b=0;b<=3;b++){
  int i = *s;
  s = &i;
 }
 *s = b;
}
\end{minted}
    \caption{A use-after-scope UB at line 8.}
    \label{fig:invalid-scope}
\end{minipage}
\end{figure}

Figure~\ref{fig:bugs_each_ub} shows the number of bugs triggered by each kind of UB. Since both \asan and \ubsan support the detection of buffer overflow, we split the found bugs into BufOverflow (\asan) and BufOverflow (\ubsan). We can observe that buffer overflow programs triggered the most number of bugs in \asan. 
Notably, \thetool detected bugs in all UB types, which highlights its strong bug detection capability and the importance of extensively testing sanitizers.
Of the 31 bugs, 29 are sanitizer FN bugs, meaning that sanitizers failed to detect UBs in them. Interestingly, we also found 2 bugs that are not sanitizer FN bugs but rather wrong reports, which means that sanitizers report a UB but with incorrect report information such as a wrong UB type warning.

\para{\tmark~Are there any false alarms by \thetool?} 
We encountered one false alarm report generated by \thetool as indicated by the ``Invalid'' row in Table~\ref{tab:bug_status}. The reported program is shown in Figure~\ref{fig:invalid-scope}. It contains a use-after-scope at line 8 because \texttt{s} points to an inner scope variable \texttt{i}. GCC \asan at -O3 can not detect it. Our crash-site mapping can verify that line 8 is still present at -O3. The GCC developers marked this report as invalid because GCC -O3 removes the for loop and moves out the inner code, which invalidates the use-after-scope UB. 
This program reveals a limitation of our crash-site mapping test oracle. Nevertheless, the significant number of reported true bugs already demonstrates its effectiveness.

\begin{table}[tp]
    \centering
    \footnotesize
    \renewcommand{\arraystretch}{1.2}
    \setlength{\tabcolsep}{3pt}
    \caption{Status of the reported bugs in GCC and LLVM.}
    \begin{tabular}{ccccccccc}
    \toprule
    \multirow{2}{*}{\textbf{Status}} & \multicolumn{2}{c}{\textbf{GCC}} && \multicolumn{3}{c}{\textbf{LLVM}} && \multirow{2}{*}{\textbf{Total}} \\
    \cmidrule{2-3}\cmidrule{5-7}
    & \asan & \ubsan && \asan & \ubsan & \msan &&\\
    \hline
    Reported & 9 & 7 && 6 & 8 & 1 && 31 \\
    Confirmed & 8 & 7 && 2 & 2 & 1 && 20 \\
    Fixed & 3 & 3 && 0 & 0 & 0 && 6 \\
    Invalid & 1 & 0 && 0 & 0 & 0 && 1 \\
    \bottomrule
    \end{tabular}
    \label{tab:bug_status}
\end{table}

\begin{figure}[tp]
    \centering
    \includegraphics[width=0.48\textwidth]{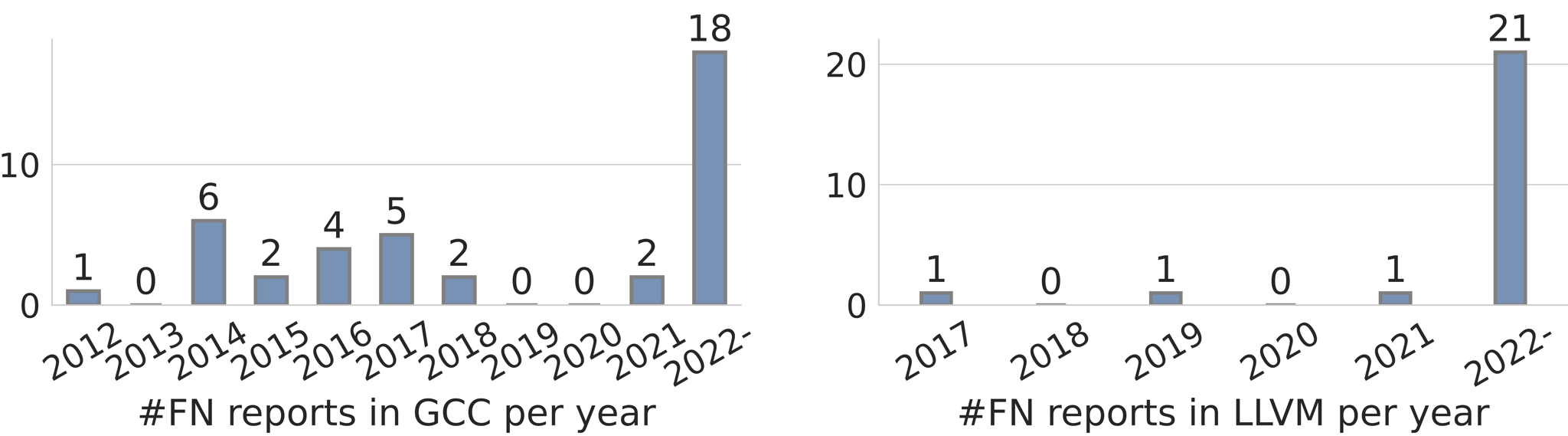}
    \caption{Number of sanitizer FN bug reports in GCC and LLVM bug trackers per year.}
    \label{fig:bugs_history}
\end{figure}

\begin{figure}[tp]
    \centering
    \includegraphics[width=0.48\textwidth]{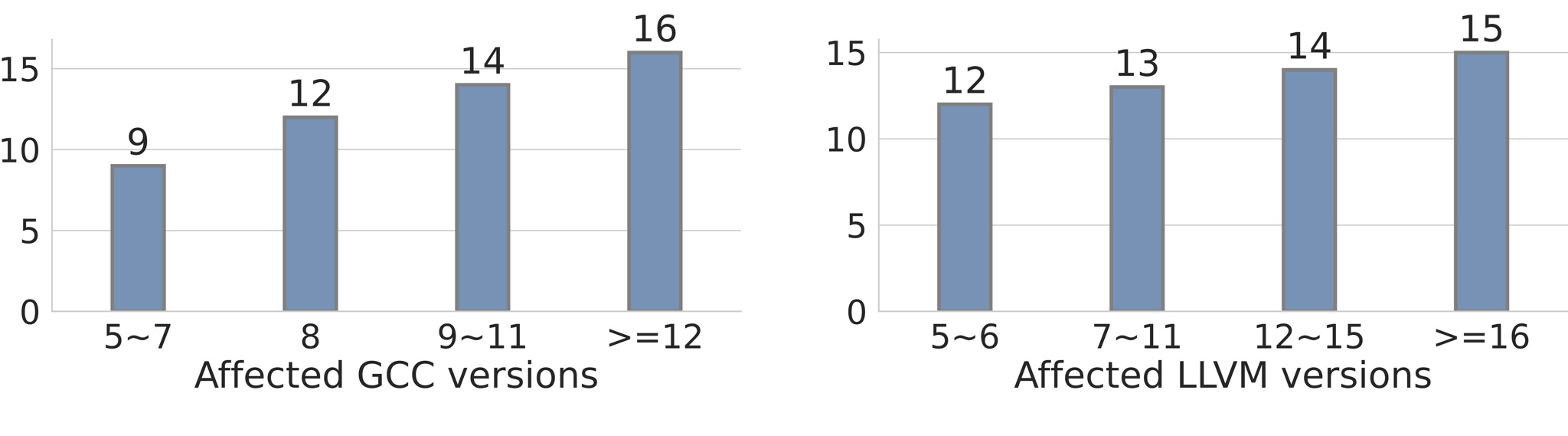}
    \caption{Stable compiler versions that are affected by the reported sanitizer FN bugs.}
    \label{fig:bugs_affected_version}
\end{figure}

\para{\tmark~How significant are the bug-finding results?}
To approach this question, we have conducted a manual analysis of all reported false negative bugs based on GCC and LLVM bug trackers of sanitizers. We choose GCC-5 (released in 2015) and LLVM-5 (released in 2017) as the earliest versions because they are the first stable versions that support sanitizers. 
The results are shown in Figure~\ref{fig:bugs_history}. 
In the past decade, there were a total of 40 false negative reports on GCC's sanitizers. Of these 40 bugs, \thetool found 16 ($40\%$). For LLVM, \thetool found 14 ($58\%$) out of the 24 bugs.
As an intermediate conclusion, \thetool has found a significant number of interesting bugs in both GCC's and LLVM's sanitizers.
To further understand the influence of our reported bugs in different stable releases of compilers, we also ran the UB programs that accompany our bug reports on all stable compiler versions. 
Figure~\ref{fig:bugs_affected_version} presents the number of sanitizer bugs that affect each stable compiler version. It indicates that \thetool can find many long-standing latent bugs, further confirming the significance of our bug-finding results.

\begin{figure}[b]
    \centering
    \includegraphics[width=0.45\textwidth]{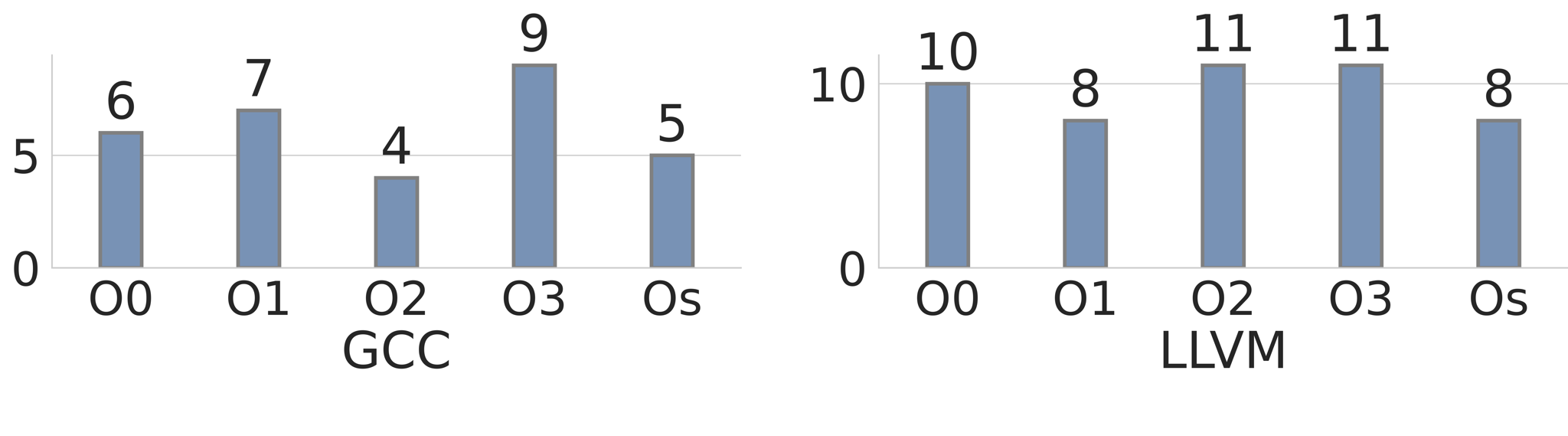}
    \caption{Affected optimization levels}
    \label{fig:bugs_affected_opt}
\end{figure}

\begin{table*}[tp]
    \centering
    \small
    \caption{The number of generated UB programs per generator. The ``No UB'' column shows the number of generated programs that do not contain UB. \thetool having ``-'' on this attribute means that all of its generated programs contain UB.}
    \renewcommand{\arraystretch}{1.2}
    \setlength{\tabcolsep}{3pt}
    \begin{tabular}{c@{\hspace{0.1cm}}cccccccccc@{\hspace{0.3cm}}c}
    \toprule
     \multirow{2}{*}{\textbf{Generator}} & \multicolumn{10}{c}{\textbf{UB}} & \multirow{2}{*}{\textbf{\makecell[c]{No \\ UB}}} \\\cmidrule{2-11}
     & \makecell[c]{Buf.Overflow\\(Pointer)} & \makecell[c]{Use After\\Free} & \makecell[c]{Use After\\Scope} & \makecell[c]{Null Ptr.\\Deref} & \makecell[c]{Integer\\Overflow} & \makecell[c]{Shift\\Overflow} & \makecell[c]{Divide\\by Zero} & \makecell[c]{Buf.Overflow\\(Array)} & \makecell[c]{Use of \\ Uninit.} & \textbf{Total} &\\
    \midrule
    \thetool & 4,213 & 3,032 & 461 & 2,082 & 408 & 287 & 329 & 2,396 & 664 & 13,872 & {-}
    \\
    MUSIC & 27 & 0 & 0 & 1 & 151 & 487 & 3 & 26 & 9 & 704 & 13,296 \\
    Csmith-NoSafe & 0 & 0 & 0 & 0 & 220 & 5,286 & 1,899 & 0 & 0 & 7,405 & 6,595 \\
    \bottomrule
    \end{tabular}
    \label{tab:gen_ub_num}
\end{table*}

\para{\tmark~Affected optimization levels.}
As shown in Figure~\ref{fig:bugs_affected_opt}, we counted the number of bugs that affect each optimization level.
The result reveals that sanitizer bugs affect all optimization levels. Testing only one of the optimization levels such as -O0 would miss many bugs that only appear at other optimization levels.
This demonstrates the usefulness of our crash-site mapping test oracle in identifying sanitizer bugs across optimization levels. 
There is no clear tendency on which optimization levels are more sensitive to sanitizer bugs.
It correlates to the fact that sanitizers and compiler optimizations work independently as having been shown in Section~\ref{sec:intro}.


\subsection{RQ2: Effectiveness of UB Program Generator}\label{sec:eval-generator}
This section provides an in-depth understanding of the effectiveness of our UB program generator. 
Although there is no other UB program generator that we could compare \thetool against, we use the following two generators as the baseline:

\begin{itemize}[labelindent=5pt, itemsep=3pt, topsep=2pt]
    \item \emph{MUSIC}~\cite{phan2018music} is a program mutator designed for mutation testing. It mutates a valid program's abstract syntax tree (AST) to generate syntactically valid mutants. By design, MUSIC may also generate UB programs as it has no guarantees regarding program semantics.
    \item \emph{Csmith-NoSafe} means that one runs Csmith with its \texttt{-no-safe-math} option. To avoid UB at runtime, Csmith utilizes many safe wrappers. For example, it changes all \texttt{x/y} to \texttt{(y==0?1:x/y)} to avoid division-by-zero. We use its \texttt{-no-safe-math} option to disable all the safe wrappers, which may introduce UB in the generated programs.
\end{itemize}

For each generator, we assess the quantity of each type of UB program that the respective generator can produce. We also equip the two baseline generators with the crash-site mapping oracle to test sanitizers.

\para{Generation quantity.}
We first use Csmith to randomly generate 1,000 seed programs. For each seed program, we use our generator to generate UB programs for every UB type that we support. 
Table~\ref{tab:gen_ub_num} details the results. The column ``Total'' shows that out of the 1,000 seed programs, \thetool generates 13,872 UB programs, averaging 14 UB programs per seed. The generated programs cover all UB types that we support. Buffer overflow takes up the most generated UB programs.
The reason is that the seeds from Csmith contain a large number of array and pointer operations, on which \thetool can generate buffer overflow programs.
Relatively fewer UB programs are generated on some of the UB types such as UseAfterScope and DividebyZero.
The main reason is that the code constructs required by them are more strict than others. For example, DividebyZero can only happen if operators ``/'' or ``\%'' are present in the live code regions. Comparatively, NullPtrDeref requires only a pointer dereference such as ``$*p$'', which apparently appears more often.

For a fair comparison to \thetool, we apply MUSIC to randomly generate 14,000 programs from the 1,000 seeds used by \thetool. Then, we utilize sanitizers\footnote{We run each program with all sanitizers. If a sanitizer reports UB on a program, we use its report to get its UB type. Note that, the programs generated by \thetool do not need such analysis because the design of \thetool allows us to know the UB type of each generated program.} to compile and analyze these programs to know if each of them contains UB. 
Table~\ref{tab:gen_ub_num} shows that there are only 704 (4\%) out of the 14,000 programs containing UB. The other 13,296 (95\%) do not contain UB.
We now use Csmith-NoSafe to generate programs. Because Csmith-NoSafe does not require a seed program, we directly use it to generate 14,000 programs. Similarly, we use sanitizers to analyze if each of the programs contains UB. From the last row in Table~\ref{tab:gen_ub_num}, we can find that around half (7,405) of the programs contain UB. This number is not as high as \thetool but already much better than MUSIC. Notably, all the UB programs are only in three types, \ie, IntegerOverflow, ShiftOverflow, and DividebyZero. This is consistent with how Csmith-NoSafe work: it removes safe wrappers around numeric operations. 
In summary, \thetool can generate the most number of UB programs and cover the most types of UB. Next, we will use MUSIC and Csmith-NoSafe as the UB generator to extensively test sanitizers.

\para{Testing sanitizers with MUSIC and Csmith-NoSafe.}
To understand if UB programs produced by the baseline generators can also find sanitizer FN bugs, we replace the generator component in \thetool with MUSIC and Csmith-NoSafe. The crash-site mapping remains unchanged to serve as the test oracle. During our testing, we let each generator generate around 1 million programs. In the end, \emph{we did not find any sanitizer FN bugs}. The failure reason for MUSIC could be that most of the generated programs did not exercise UB. For Csmith-NoSafe, its failure is mainly due to (1) the narrow range of UB types it can generate, and (2) unlike our generator, it typically introduces multiple UB in a program, which makes it hard to discover missed sanitizer reports.

\para{Testing sanitizers with the existing UB test suite.}
The Juliet test suite~\cite{juliet} released by NIST consists of a collection of UB programs. It is by far the most comprehensive test suite for UB detectors. To understand if UB programs from the existing test suite can find sanitizer bugs, we select all the 16,344 UB programs from the Juliet test suite that are detectable by sanitizers. Instead of using a generator, we directly use all the UB programs from the test suite as the source of programs. Our results show that \emph{none of the UB programs from the Juliet test suite can find sanitizer FN bugs.} This further confirms the necessity of a UB program generator like ours.



\subsection{RQ3: Effectiveness of Crash-Site Mapping}\label{sec:eval-oracle}
For each generated UB program, we apply differential testing to find discrepancies across compilers.
We then use our crash-site mapping to determine if a discrepancy is caused by a sanitizer FN bug or merely compiler optimizations.
For the 13,872 UB programs generated from Section~\ref{sec:eval-generator}, we run all the sanitizers specified in the evaluation setup (Section~\ref{sec:eval-setup}) to select programs that cause discrepant sanitizer reports. This results in a total of 6,567 selected programs, nearly half of the generated UB programs. The substantial number of discrepancy-causing programs highlights (1) the exceptional quality of our generated UB programs, and (2) without our crash-site mapping, discerning real sanitizer bug-caused discrepancies from the 6,567 discrepancies would be practically infeasible.
To evaluate the effectiveness of our crash-site mapping test oracle, we measure its \emph{precision} and \emph{recall}.

\para{Precision:} \emph{Out of all selected discrepancies, how many are truly caused by sanitizer bugs?}
Out of the \textbf{6,567} discrepancies, our crash-site mapping selected \textbf{58} and dropped the rest \textbf{6,505} as invalid. For each of the selected discrepancies, we manually verify if it is caused by compiler optimizations. \emph{Our manual analysis found that all discrepancies selected by crash-site mapping are due to sanitizer bugs, which means that our crash-site mapping achieves perfect precision.} Although we have analyzed an invalid report by \thetool in Section~\ref{sec:bug-finding}, it does not appear in our quantitative evaluation. Thus, we may conclude that our crash-site mapping has a high precision.

\para{Recall:} \emph{Out of all sanitizer bug-caused discrepancies, how many are selected?}
This measures if our crash-site mapping will miss interesting discrepancies. Ideally, we should analyze all the dropped discrepancies to verify if any of them are due to sanitizer bugs. However, this requires a manual analysis of 7,966 discrepancies. To reduce the cost, we randomly sampled 200 dropped discrepancies by the crash-site mapping and then manually analyzed each of them.
Perhaps surprisingly, after our analysis, we found that none of the dropped discrepancies were caused by sanitizer bugs. In other words, \emph{our crash-site mapping achieves 100\% recall} on these samples.
Since our evaluation is on sampled data, it is not complete, but it does suggest that crash-site mapping has a high recall.  

\revision{
\para{Soundness of Crash-Site Mapping:}
As defined in Definition~\ref{def:crashsite}, the \emph{crash site} is associated with the source location of the last executed instruction. The soundness of \emph{crash-site mapping} largely depends on a reliable mapping between instructions and source locations. In our implementation, we enable \texttt{-g} option for all compilations, which enriches the produced binaries with debugging meta-data. These meta-data can then be utilized by a debugger to obtain the source location, \ie, \emph{(line number, offset)}, of each instruction. 
Although compiler optimizations can remove instructions with their meta-data, it will not cause the soundness problem in \emph{crash-site mapping} because this resides in the scope discussed in Challenge 2 in \S\ref{sec:intro}.
Unfortunately, a recent study~\cite{theodoridis2022finding} has confirmed that bugs in compilers may lead to incorrect debugging meta-data. Buggy meta-data can theoretically cause unsound or incorrect \emph{crash-site mapping} results. For instance, \emph{crash-site mapping} can incorrectly flag the existence of an eliminated crash-site, and thus generate false positive reports.
During our extensive testing period, we did not observe any false positive reports though. We believe such compiler bugs to be rare in our testing scenario.
Handling buggy debugging meta-data is an orthogonal research program and we assume always correct meta-data in this work.
}

\begin{table}[tp]
    \centering
    \footnotesize
    \renewcommand{\arraystretch}{1.2}
    \setlength{\tabcolsep}{3pt}
    \caption{Line coverage (LC), function coverage (FC), and branch coverage (BC) of GCC and LLVM.}
    \begin{tabular}{lccclccc}
    \toprule
    & \multicolumn{3}{c}{\textbf{GCC}} && \multicolumn{3}{c}{\textbf{LLVM}}\\
    \cmidrule{2-4}\cmidrule{6-8}
    & LC & FC & BC && LC & FC & BC\\
    \midrule
    Seeds & 63.1\% & 65.5\% & 49.4\% && 30.4\% & 38.2\% & 23.3\% \\
    MUSIC & 63.1\% & 65.5\% & 49.4\% && 30.5\% & 38.2\% & 23.4\% \\
    Csmith-NoSafe & 63.6\% & 65.5\% & 50.1\% && 32.5\% & 40.2\% & 24.8\% \\
    \thetool & {63.7\%} & {65.5\%} & {50.8\%} && {31.8\%} & {39.3\%} & {24.3\%} \\
    \bottomrule
    \end{tabular}
    \label{tab:eval-cov}
\end{table}

\subsection{RQ4: Code Coverage}
We utilized Gcov and only instrumented sanitizer-related files to collect coverage in both GCC and LLVM. We used the generated programs from Section~\ref{sec:eval-generator} to profile coverage. 
Table~\ref{tab:eval-cov} summarizes our results.
In all cases, compared to the seed programs, all generators lead to a moderate coverage improvement, with \thetool and Csmith-NoSafe showing the largest increase on GCC and LLVM, respectively.

\begin{figure*}[tp]
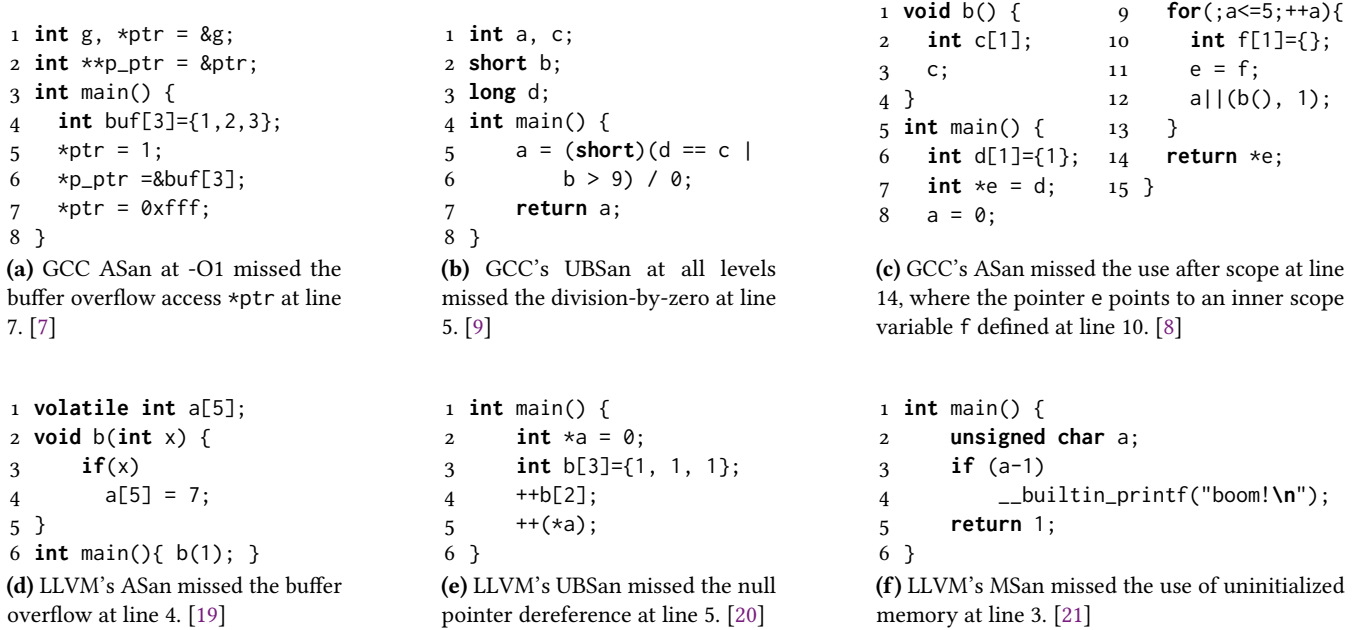

    \captionsetup[subfigure]{aboveskip=1pt}
    \vspace{20pt}

    \begin{subfigure}[b]{.25\textwidth}
\begin{minted}[linenos,xleftmargin=1em,fontsize=\small]{C}
int g, *ptr = &g;
int **p_ptr = &ptr;
int main() { 
  int buf[3]={1,2,3};
  *ptr = 1;
  *p_ptr =&buf[3];
  *ptr = 0xfff;
}
\end{minted}
    \caption[1]{GCC \asan at -O1 missed the buffer overflow access \texttt{*ptr} at line 7.~\cite{gcc106558}}
    \label{fig:bug-gcc-asan-bufferoverflow}
    \end{subfigure}
\hfill
    \begin{subfigure}[b]{.25\textwidth}
\begin{minted}[linenos,xleftmargin=1em,fontsize=\small]{C}
int a, c;
short b;
long d;
int main() { 
    a = (short)(d == c |
        b > 9) / 0; 
    return a;
}
\end{minted}
    \caption{GCC's UBSan at all levels missed the division-by-zero at line 5.~\cite{gcc109151}}
    \label{fig:bug-gcc-ubsan-dividebyzero}
    \end{subfigure}
\hfill
    \begin{subfigure}[b]{.35\textwidth}
\begin{multicols}{2}
\setlength{\columnsep}{1cm}
\begin{minted}[linenos,xleftmargin=1em,fontsize=\small]{C}
void b() {
  int c[1];
  c;
}
int main() {
  int d[1]={1};
  int *e = d;
  a = 0;
\end{minted}
\columnbreak
\begin{minted}[firstnumber=9,linenos,xleftmargin=0em,fontsize=\small]{C}
  for(;a<=5;++a){
    int f[1]={};
    e = f;
    a||(b(), 1);
  }
  return *e;
}
\end{minted}
\end{multicols}
    \caption{GCC's ASan missed the use after scope at line 14, where the pointer \texttt{e} points to an inner scope variable \texttt{f} defined at line 10.~\cite{gcc108085}}
    \label{fig:bug-gcc-asan-useafterscope}
    \end{subfigure}

    \vspace{20pt}
    \begin{subfigure}[b]{.25\textwidth}
\begin{minted}[linenos,xleftmargin=1em,fontsize=\small]{C}
volatile int a[5];
void b(int x) {
    if(x)
      a[5] = 7;
}
int main(){ b(1); }
\end{minted}
    \caption{LLVM's ASan missed the buffer overflow at line 4.~\cite{llvm55189}}
    \label{fig:bug-llvm-asan-bufferoverflow}
    \end{subfigure}
\hfill
    \begin{subfigure}[b]{.25\textwidth}
\begin{minted}[linenos,xleftmargin=1em,fontsize=\small]{C}
int main() {
    int *a = 0;
    int b[3]={1, 1, 1};
    ++b[2];
    ++(*a);
}
\end{minted}
    \caption{LLVM's UBSan missed the null pointer dereference at line 5.~\cite{llvm60236}}
    \label{fig:bug-llvm-ubsan-nullptrderef}
    \end{subfigure}
\hfill
    \begin{subfigure}[b]{.35\textwidth}
\begin{minted}[linenos,xleftmargin=1em,fontsize=\small]{C}
int main() {
    unsigned char a;
    if (a-1)
        __builtin_printf("boom!\n");
    return 1;
}
\end{minted}
    \caption{LLVM's MSan missed the use of uninitialized memory at line 3.~\cite{llvm61982}}
    \label{fig:bug-llvm-msan-useuninit}
    \end{subfigure}

\caption{Sample UB programs that trigger sanitizer FN bugs.}
\end{figure*}

\subsection{Case Study}

\begin{table}[tp]
    \centering
    \footnotesize
    \renewcommand{\arraystretch}{1.2}
    \setlength{\tabcolsep}{10pt}
    \caption{\revision{Bug category according to root cause analysis.}}
    \revision{
    \begin{tabular}{lcc}
    \toprule
    \textbf{Category} & \textbf{GCC} & \textbf{LLVM} \\
    \midrule
    No Sanitizer Check & 2 & 2 \\
    Incorrect Sanitizer Optimization & 5 & 3 \\
    Wrong Red-Zone Buffer & 1 & 1 \\
    Incorrect Sanitizer Check & 2 & 7 \\
    Incorrect Expression Folding/Shorten & 4 & 1 \\
    Incorrect Operation Handling & 0 & 1 \\
    Wrong Line Information & 2 & 0 \\
    \bottomrule
    \end{tabular}
    }
    \label{tab:bug-category}
\end{table}

\revision{
In order to understand the reason why sanitizers make mistakes, we categorize all bugs according to their root causes. The categorization is based on both our manual analysis and developers' feedback. Table~\ref{tab:bug-category} shows the result.
Both GCC and LLVM make some common mistakes. For example, their sanitizer implementations may conduct ``Incorrect Sanitizer Optimization'' causing valid sanitizer checks to be removed.
We discuss a selection of representative bugs in each bug category. 
}

\para{Figure~\ref{fig:bug-gcc-asan-bufferoverflow}:} \revision{(\emph{No Sanitizer Check})}
This program contains an overflowed memory access at line 7.
Since \texttt{p\_ptr} initially points to pointer \texttt{ptr} at line 2, \texttt{ptr} will point to the overflowed address \texttt{\&buf[3]} after line 6. Therefore, a stack-buffer-overflow occurs at line 7, and then the value \texttt{0xfff} is written to \texttt{buf[3]}.
However, due to a sanitizer instrumentation bug, GCC \asan at -O2
fails to insert the check for the validity of \texttt{*ptr} at line 7, and thus cannot report it. 
This bug affects GCC trunk and has been fixed.

\para{Figure~\ref{fig:bug-gcc-ubsan-dividebyzero}:} \revision{(\emph{Incorrect Expression Folding/Shorten})}
This program reveals a long latent bug in GCC \ubsan, which fails to report the DivisionbyZero UB at line 6. The root cause is that \ubsan only cares about integer operands, but not booleans. However, although \texttt{(d==c|b>9)} is boolean, it gets widened to \texttt{short}. GCC \ubsan incorrectly handles this case and thus misses the UB. This bug exists since the introduction of UBSan in GCC.

\para{Figure~\ref{fig:bug-gcc-asan-useafterscope}:} \revision{(\emph{Incorrect Sanitizer Optimization})}
This program contains a UseAfterScope UB at line 14, where \texttt{e} points to an inner scope variable \texttt{f}. GCC \asan fails to report this bug at -O3. In fact, GCC \asan initially indeed inserts a scope check for \texttt{f} at line 10, but another sanitizer analysis module removes this check when exiting the loop. 

\para{Figure~\ref{fig:bug-llvm-asan-bufferoverflow}:} \revision{(\emph{Wrong Red-Zone Buffer})}
This program has an overflowed array access at line 4, where the array \texttt{a} is of length \texttt{5}. LLVM \asan incorrectly marks the overflow access as within the scope of array padding while in fact, it is not. This bug reveals a fundamental problem with \asan handling of global arrays. It affects all LLVM versions at all optimization levels.

\para{Figure~\ref{fig:bug-llvm-ubsan-nullptrderef}:} \revision{(\emph{Incorrect Sanitizer Check})}
This program contains a NullPointerDereference UB at line 5, where the pointer \texttt{a} is NULL and the program tries to increment it. LLVM \ubsan does not report this bug because the null pointer check is not placed before the increment operation. The developer believes that the \texttt{++} operator misleads 
\ubsan's internal logic because if we replace \texttt{++(*a)} with \texttt{*a += 1}, \ubsan would work again.

\para{Figure~\ref{fig:bug-llvm-msan-useuninit}:} \revision{(\emph{Incorrect Operation Handling})}
The \texttt{if} branch in this program can be taken differently depending on the value of uninitialized variable \texttt{a}. LLVM \msan incorrectly handles the subtraction and thinks that the value of \texttt{(a-1)} is fully determined. The LLVM developers have confirmed this bug and are working on a fix.

\onecolumn \begin{multicols}{2}
    
\subsection{Discussion on Approach Generality}



Despite sanitizers are the most popular UB detectors, there are many other dynamic and static UB detection tools. Dynamic tools such as Dr. Memory~\cite{bruening2011practical} and Valgrind~\cite{nethercote2007valgrind} can detect memory errors including buffer overflows, use of uninitialized memory, improper free, \etc Static tools such as CppCheck~\cite{cppcheck} and Infer~\cite{infer} can detect null pointer dereferences, integer overflows, \etc 
In principle, our approach can also be used to test these detectors. We currently focus on sanitizers because they have a wider real-world impact, especially in the area of fuzzing. Our evaluation results on testing sanitizers have already confirmed the significant UB program generation and bug-finding capability of our tool. Extending our testing scope to other detectors would be an interesting application of our approach and help solidify these additional tools.



\section{Related Work}

\para{Compiler Testing.}
Finding compiler bugs has been extensively studied; significant research effort has been devoted to testing various compiler functionalities.
Csmith~\cite{li2023finding} is the most popular program generator for C and has found hundreds of compiler crashes and correctness bugs. Csmith-generated programs are guaranteed to be free of undefined behavior.
Instead of generation, Equivalent Modulo Input (EMI)~\cite{le2014compiler} is proposed to mutate/transform a seed program while preserving its semantics under the same input. 
EMI can be implemented by deleting dead statements~\cite{le2014compiler}, inserting new code in dead regions~\cite{le2015finding}, or synthesizing equivalent code in live regions~\cite{sun2016finding}.
Together with Csmith, EMI has found thousands of compiler optimization bugs.
YARPGen~\cite{livinskii2020random} is another C/C++ program generator that aims to test scalar optimizations in compilers.

In addition to optimization correctness, other issues in compilers such as incorrect debug information~\cite{li2020debug,wang2023compilation} and missed optimizations~\cite{theodoridis2022finding} have also been studied.
Li~\etal~\cite{li2020debug} construct the so-called actionable programs to validate the debug information generated for optimized code.
Dfusor~\cite{wang2023compilation} transforms a seed program into multiple variants and then uses them to find debug information inconsistencies.
Dead~\cite{theodoridis2022finding} injects markers into dead regions of a program to find missed dead code eliminations in optimizing compilers. 


\para{Sanitization.}
\asan and \msan use shadow memory to record and check the safety of each memory access. 
Runtime checks are inserted around memory accesses during the compilation of a program.
Similarly, \ubsan uses tailored checks for different UBs such as overflow checks for additions and null pointer checks for pointer dereferences.
These checks will inevitably increase a program's runtime overhead. Many approaches have been proposed to reduce the overhead by removing redundant checks~\cite{zhang2021sanrazor}, optimizing checks~\cite{zhang2022debloating}, or applying checks to only a subset of the original code~\cite{wagner2015high,lettner2018partisan}.
These optimizations are meaningful in improving sanitizers' practical utility. \thetool can also be used to validate their implementations once they are integrated into mainstream compilers.

\section{Conclusion}

We have presented a novel framework for testing sanitizer implementations. 
We have introduced a UB program generator that generates UB programs from a seed program via shadow statement insertion. Based on this generator, we have employed differential testing across multiple compilers to test sanitizers. To filter out discrepancies caused by compiler optimizations, we have designed a new test oracle, crash-site mapping, that is capable of accurately identifying true sanitizer bugs. \thetool, our implementation of the testing framework, has discovered 31 bugs in \asan, \ubsan, and \msan from both GCC and LLVM.
Our work represents a promising, initial step toward comprehensive validations of sanitizer implementations, and highlights the importance of this problem.

\section{Acknowledgments}

We thank the anonymous ASPLOS reviewers for their valuable feedback.
Our special thanks go to the GCC and LLVM developers for useful information and for addressing our bug reports. This work was partially supported by a Meta Security Research RFP award.

\appendix
\section{Artifact Appendix}
\subsection{Abstract}

The artifact contains the code and datasets we used for our experiments, as well as scripts to generate the numbers and tables of our evaluation.
Specifically, it includes (a) links and bug-triggering test cases of each reported bug; (b) 1,000 Csmith seed programs used for evaluation; (c) scripts for generating UB programs with \thetool, MUSIC, and Csmith-NoSafe; (d) scripts for reporting coverage achieved by each apprach; and (e) detailed instruction documentation for using \thetool.
Everything is packaged and pre-built as a docker image. A standard X86 Linux machine running docker is necessary
to evaluate this artifact.

\subsection{Artifact Check-List (Meta-Information)}

{\small
\begin{itemize}
  \item {\bf Run-time environment: } Linux
  \item {\bf Hardware: } X86
  \item {\bf Output: } Statistics of CompDiff detection results on the Juliet testsuite and 23 real-world programs.
  \item {\bf How much disk space required (approximately)?: } 40GB
  \item {\bf How much time is needed to prepare workflow (approximately)?: }10-20 minutes to download and import the docker image.
  \item {\bf How much time is needed to complete experiments (approximately)?: } 20 hours
  \item {\bf Publicly available?: } Yes
  \item {\bf Code licenses (if publicly available)?: }Apache 2.0
  \item {\bf Archived (provide DOI)?: } Yes
\end{itemize}
}


\subsection{Description}

\subsubsection{How to access}

The artifact can be downloaded from the following link:\\
\url{https://doi.org/10.5281/zenodo.8406414}

\subsubsection{Hardware dependencies}

A standard X86 machine.
\subsubsection{Software dependencies}
Docker

\subsection{Installation}
tar xf compdiff-asplos23-ae.tar.gz\\
cat compdiff-asplos23-image.tar | docker import - compdiff\_ae

\subsection{Experiment Workflow}
\begin{enumerate}
    \item Read the documentation.
    \item Start the docker container as instructed.
    \item Check bug reports.
    \item Run \thetool, MUSIC, and Csmith-NoSafe to generate UB programs.
    \item Collect coverage information for each approach.
    
\end{enumerate}
\subsection{Evaluation and Expected Results}

We provide data and scripts to generate all the evaluation results in Section 4.
Specifically, Tables 3, 4, and 5 are reproduced.

\bibliographystyle{ACM-Reference-Format}
\bibliography{references}
\end{multicols}
\end{document}